\documentclass[11pt, a4paper, pdftex]{article}
\usepackage[utf8]{inputenc}		
\usepackage[T1]{fontenc}

\usepackage{amsmath}
\usepackage{amssymb}
\usepackage{setspace}
\usepackage{latexsym}
\usepackage{fancyhdr}
\usepackage{lastpage}
\usepackage{times}

\usepackage[pdftex]{graphicx}

\usepackage[english]{babel}

\usepackage{authblk}

\graphicspath{
	{my_figures/}
}

\newcommand{\ji}{\mathrm{j}}

\newcommand{\pa}{\paragraph{}}
\newcommand{\di}{\displaystyle}

\newcommand{\dd}{\frac{d}{2}}
\newcommand{\sdd}{\frac{d^2}{4}}

\newcommand{\s}{\underline{s}}
	
\def\og{\leavevmode\raise.3ex\hbox{$\scriptscriptstyle\langle\!\langle$~}}
\def\fg{\leavevmode\raise.3ex\hbox{$\scriptscriptstyle\rangle\!\rangle$}}

\begin{document}

\title{Listening broadband physical model for microphones: a first step}
\author[1, 2]{Laurent Millot \thanks{Electronic address: l.millot@ens-louis-lumiere.fr; corresponding author}}
\author[2]{Antoine Valette}
\author[2]{Manuel Lopes}
\author[1, 2]{Gérard Pelé}
\author[2]{Mohammed Elliq}
\author[1, 2]{Dominique Lambert}
\affil[1]{Institut d'esthétique, des arts et technologies - UMR 8153 (CNRS/Université Paris 1/MENRT), 
Université Paris 1, 47 rue des bergers, Paris, 75015, France}
\affil[2]{ENS Louis-Lumière, Noisy-le-Grand, 7 all\'{e}e du promontoire, 93160, France}

\date{15/03/2006}

\maketitle

	\begin{abstract}
We will present a first step in design of a broadband physical model for microphones. Within the proposed model, classical directivity patterns (omnidirectional, bidirectional and cardioids family) are refound as limit cases: monochromatic excitation, low frequency and far-field approximation. Monophonic pieces of music are used as sources for the model so we can listen the simulation of the associated recorded sound field in realtime thanks to a Max/MSP application. Listening and subbands analysis show that the directivity is a function of frequential subband and source location. This model also exhibits  an interesting proximity effect. Audio demonstrations will be given.

Paper 6638 presented at the 120th Convention of the Audio Engineering Society, Paris, 2006
	\end{abstract}

\section{Introduction}
\pa When one need to qualify the behavior of a microphone or build 
 a multichannel recording setup \cite{Williams:04, Remy:04, Braasch:05, Martin:05}, the same classical directivity function $\di D(\theta)$ is commonly used. This directivity function is given by:
\begin{equation}
D(\theta)=m + (1-m).\cos\theta
\label{eq:Dir}
\end{equation}
where 
\begin{itemize}
\item $\theta$ is the incidence angle, 
\item $m$ a coefficient varying from 0 to 1 according to the desired directivity ($m=1$: omnidirectional, $m=0$: bidirectional, $m=0.5$: cardioid, ...). 
\end{itemize}

\pa And it was the case for the stereophonic recording simulator developed as a part of their master thesis by Carr-Brown, Colcy and Delatte \cite{CCD:04} and used for the study of the composition of auditory scenes \cite{CCD:04, Millot:05}. The microphone model used for this works was simply built by adding a spherical wave-propagation filtering to the effect of the directivity function.

\pa Despite its simplicity, this model assumes that the directivity is the same for the whole audio spectrum which is not the case when considering the data provided by the microphones manufacturers \cite{Shure, Neumann, DPA, Behringer} as the directivity diagrams exhibit a behavior which can significantly vary with frequency.

\pa Moreover, this model does not permit to refind a proximity effect as it is derived in a monochromatic, low frequency and far-field approximation. 

\pa Considering  these limitations and troubles between theory and experimental data, we have been working on derivation of  a physical model able to exhibit a behavior  as real as possible. In this article, we present   a model:
\begin{itemize}
\item exhibiting a behavior varying with the frequency and the source location (distance and incidence);  
\item exhibiting  also a proximity effect;
\item which can be used for numerical simulations in realtime, using broadband monophonic sound files as source, one can listen in order to study the behavior of a virtual approximation of a microphone.
\end{itemize}

\pa We also review the potential ways to improve this first model and extend this simulation of recording to multi-microphones setups.  In this paper, we had prefer to concentrate our attention on the theory, in order to precise every needed elements, as 
we think the best appreciation tools for such a work is the listening trial. But, we will discuss some clues to build some new tools more close to the human perception and to the "everyday" behavior of a microphone:  use of complexe real sounds rather than sinusoids or noise.

\pa This first model has been proposed by Millot while the Max/MSP application has been first developed by Valette as a part of its master thesis \cite{Valette:05}.

\section{Physical model for the captured sound field}
\subsection{Assumptions}
\paragraph{High level assumptions: }Within this article, we intend to derive a physical model of the captured acoustical sound field. This restriction means that we do not consider the presence of the microphone in the sound field. In fact, we synthesize the captured signal as a linear combination of several acoustical sound fields. Even if quite strange, this  assumption is absolutely coherent with the derivation of the classical directivity function. 

\pa By acoustical sound field, we mean the theoretical pressure field
created in the whole considered space by the presence of a given source at a given location.

\pa In order to derive the model for the captured sound field, we consider the resultant directional sound field as the combination of two different components:
\begin{itemize}
\item an omnidirectional acoustical sound field associated with a point chosen as center of the "fictitious microphone" location;
\item a bidirectional sound field which is built from the difference of two punctual acoustical sound fields, giving a so-called dipolar sound field, associated with two points symmetrically spaced around the center of the "fictitious" microphone.
\end{itemize}

\pa One may note that we search to synthesize an omnidirectional and a bidirectional sound fields to be able to refind the classical directivity (based on the combination of omnidirectional and bidirectional directivities) as a limit case: the monochromatic, low frequency and far-field approximation.

\paragraph{Low level assumptions: }We consider a punctual source where the emitted signal is a monophonic sound file within the Max/MSP realtime application.

\pa Each of the acoustical sound fields used to synthesize the directional resultant sound field are also punctual.

\pa From the source location to each of three needed "captation" locations we consider the propagation of a spherical wave without atmospheric absorption, as we consider this work as a first step.

\pa We consider that the source and the three captation points are coplanar in the following to keep the simplicity of the presentation  and because the classical directivity function corresponds the coplanar case. But, the extension to a non coplanar situation is rather easy to perform and, moreover, to program: one just need to introduce a third coordinate in the calculation of the distances between the source and the three respective captation points.

\pa We choose the center of the fictitious microphone as origin for the coordinates because such a choice will simplify the mathematical expressions and will permit a modular design. Indeed, the development of a "multi-microphones" version of the simulator will be easier if we re-use as many "microphone" modules as needed, each of them using a local referential whose origin is the center of the considered "microphone". 

\pa We do not take into account a room effect for this first model, which means that we work within the free-field assumption, because we consider that the main difficulties are related to the description of the acoustical phenomena occurring between the source and the captation points for the so-called "direct sound". 

\pa As we also need a digital version of the model to program the realtime simulator, we note $F_s$ the sampling rate which will be given by the sound file used as source. In the first Max/MSP prototype, the sampling rate is 44.1 kHz and the quantization depth is chosen equal to 16 bits. 

\section{Derivation of the physical model}
\subsection{Omnidirectional sound field}
\pa We note $s(t)$ the analog signal associated with the source, and $s[n]$ its digital version.

\pa If the distance between the source and the center point is noted $r$, the acoustical pressure founded in the center point, $s_{omni}(r,\theta, t)$ is then:
\begin{equation}
s_{omni}(r,\theta,t)=\frac{1}{r}.s(t-\frac{r}{c_0})
\label{eq:Omni}
\end{equation}
where $c_0$ corresponds to the speed of sound.

\pa As the signal found in the center point does not  depend on the incidence of the source, this acoustical sound field is omnidirectional. 

\pa The signal $s_{omni}(r,\theta,t)$ is an attenuated version and delayed version of the source signal $s(t)$: geometrical attenuation associated with the $\frac{1}{r}$ factor and delay of $\tau_0=\frac{r}{c_0}$ seconds. 

\pa So, to get the digital version of the omnidirectional acoustical sound field, we just need to calculate the digital delay $D_0$, given in samples,  associated with the delay $\tau_0$, given in seconds. In fact, we can notice that $\tau_0$ and $D_0$ are linked by the following relation:
\begin{equation}
D_0=F_s.\tau_0=F_s.\frac{r}{c_0}.
\label{eq:Del0}
\end{equation}

\pa The digital version of the signal captured at the center point is then:
\begin{equation}
s_{omni}[n]=\frac{1}{r}.s[n-D_0]
\label{eq:SigOmni}
\end{equation}
which means that $\di s_{omni}[n]$ corresponds to the filtered version of the source signal $s[n]$ resulting from the application 
of the omnidirectional acoustical sound field digital filter: 
\begin{equation}
H_{omni}[z]=\frac{1}{r}.z^{-D_0}.
\label{eq:FilOmni}
\end{equation}

\pa One can note that this filter corresponds to the series association of an amplificator of gain $\frac{1}{r}$ and of a pure delay $z^{-D_0}$ as illustrated in Fig~1.
 
\begin{figure}[h!t]
\begin{center}
\includegraphics[height=2cm]{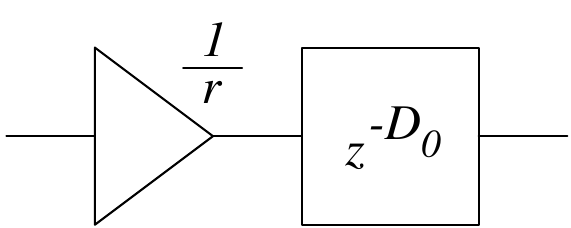}
\end{center}
\caption[1]{Digital filter to apply to the source signal to derive an omnidirectional acoustical sound field.}
\end{figure}

\subsection{Bidirectional sound field}
\paragraph{Captured dipolar sound field : }The dipolar acoustical sound field is given by the difference of the acoustical sound fields associated with two points, symmetrical rather to the center of the "fictitious" microphone and spaced by a distance $d$, as illustrated by Fig~2.

\begin{figure}[h!t]
\begin{center}
\includegraphics[height=5cm]{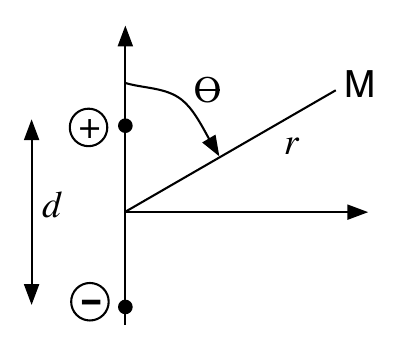}
\end{center}
\caption[2]{Definition of the geometrical configuration used to build the dipolar acoustical sound field.}
\end{figure}

\pa We note, $r_1$ and $r_2$ the respective distance between the source and both captation points and consider that the acoustical sound field related to the second point is subtracted from the sound field associated with the first one. In such conditions, the dipolar sound field $s_{dipo}(r,\theta, t)$ is given by the subtraction of both omnidirectional acoustical sound fields:
\begin{equation}
s_{dipo}(r,\theta, t)=\frac{1}{r_1}.s(t-\frac{r_1}{c_0})-\frac{1}{r_2}.s(t-\frac{r_2}{c_0}).
\label{eq:ChpDipo}
\end{equation}

By introducing the digital delays $D_1$ and $D_2$ related to $\tau_1=\frac{r_1}{c_0}$ and  $\tau_2=\frac{r_2}{c_0}$ by:
\begin{equation}
D_i=F_s.\tau_i=F_s.\frac{r_i}{c_0}\;\;\;\;\;\;\; i=1,2
\label{eq:Del0}
\end{equation}
we observe that the digital version of the dipolar sound field is then given by
\begin{equation}
s_{dipo}[n]=\frac{1}{r_1}.s[n-D_1]-\frac{1}{r_2}.s[n-D_2]
\label{eq:SiDipo}
\end{equation}
which corresponds to a version of the source signal $s[n]$ filtered by:
\begin{equation}
H_{dipo}[z]=\frac{1}{r_1}.z^{-D_1}-\frac{1}{r_2}.z^{-D_2}.
\label{eq:SiDipo}
\end{equation}

\pa We introduce the following conventions:
\begin{itemize}
\item center of the first omnidirectional sound field $O_1(\dd,\;0)$;

\item center of the second omnidirectional sound field $O_2(-\dd,\;0)$ (to be subtracted from the first one);

\item center of the "fictitious" microphone $O(0,\;0)$;

\item source location $M(r\cos\theta,\; r\sin\theta)$;

\item distance source/ point \#1 : $r_1=O_1M$ which means $\di r_1=\sqrt{r^2+\sdd-rd\cos\theta}$;

\item distance source/ point \#2 : $r_2=O_2M$ which means $\di r_2=\sqrt{r^2+\sdd+rd\cos\theta}$.
\end{itemize}

\paragraph{Captured bidirectional sound field : }To derive the expression of the bidirectional sound field, and refind the classical directivity function as a limit case, we need to consider the monochromatic source case. And, we then search the useful modifications of the dipolar sound field to find the bidirectional sound field by considering the analog case. By the end of the work of Valette \cite{Valette:05}, we have found that such a method has also been used by Cotterell \cite{Cotterell:02} but considering that the thickness of the membrane is negligible and then using a pressure gradient component rather than our dipolar sound field. The digital version of the model will be derived easily, from the analog case, at the end of the calculation.

\pa In order to simplify the calculation, and as we consider the case of monochromatic excitation,we use the phasor notation:
\begin{equation}
s(r,\theta,t)=\Re\Big(\s(r, \theta,t)\Big)=\Re\Big(\frac{A}{r}.e^{\ji (\omega t  -kr)}\Big)
\label{eq:Modele}
\end{equation}
where $\Re(\s)$ corresponds to the real part of the phasor $\s$.

\pa With the use of the phasor, the dipolar sound field is then given by:
\begin{equation}
\s_{dipo}(r,\theta,t)=\frac{A}{r_1}.e^{\di\ji \omega (t-\frac{r_1}{c_0})}-\frac{A}{r_2}.e^{\di\ji \omega (t-\frac{r_2}{c_0})}.
\label{eq:DipChp2}
\end{equation}

\pa Considering the \textbf{geometrical} far-field location of the 
source, related to the \textbf{geometrical condition} $d\ll r$, we can introduce the following approximations:
\begin{itemize}
\item $\di r_1 \approx r -\frac{d}{2}\cos\theta \approx  r$;
\item $\di r_2 \approx r +\frac{d}{2}\cos\theta \approx  r$. 
\end{itemize}   

\pa With these approximations, we can rewrite the dipolar sound field as:
\begin{equation}
\s_{dipo}(r,\theta,t)\approx \frac{A}{r}.e^{\;\di\ji \omega (t- \frac{r}{c_0})}.
\Big(e^{\di\ji \frac{\omega d.\cos\theta}{ 2c_0}} 
-  e^{\di -\ji \frac{\omega d.\cos\theta}{ 2c_0}} \Big)
\label{eq:DipChp3}
\end{equation}
or, with $\di\s_{omni}(r,\theta,t)=\frac{A}{r}.e^{\ji \omega (t-\frac{r}{c_0})}$,
\begin{equation}
\s_{dipo}(r,\theta,t)\approx 
2\ji \sin\Big( \frac{\omega d.\cos\theta}{2c_0}\Big).
\s_{omni}(r,\theta,t).
\label{eq:DipChp4}
\end{equation}

\pa To point out the bidirectional characteristic, we also need to consider that the condition $f\ll \frac{c_0}{\pi d}$ which corresponds to a low-frequency assumption for the excitation. Within this new approximation, the dipolar sound field becomes:
\begin{equation}
\s_{dipo}(r,\theta,t)\approx 
\ji \omega \frac{d}{c_0}.\cos\theta.
\s_{omni}(r,\theta,t).
\label{eq:DipChp5}
\end{equation}

\pa As the bidirectional acoustical sound field is given by relation 
\begin{equation}
\s_{bidi}(r,\theta,t)= \cos\theta.
\s_{omni}(r,\theta,t)
\label{eq:DipChp6}
\end{equation}
we conclude that the relation between bidirectional and dipolar sound fields is:
\begin{equation}
\s_{dipo}(r,\theta,t)\approx 
\ji \omega \frac{d}{c_0}.\s_{bidi}(r,\theta,t).
\label{eq:DipChp7}
\end{equation}

\pa Then, in the monochromatic, low-frequency and far-field geometrical assumptions, the bidirectional sound field $s_{bidi}(r,\theta,t)$ corresponds to the filtering of the dipolar sound field $s_{dipo}(r,\theta,t)$ by the analog filter of transfert function:
\begin{equation}
H(\omega)=\frac{c_0}{d}.\frac{1}{\ji \omega}
\label{eq:FilDipo}
\end{equation}
which represents the series association of an amplificator of gain $\frac{c_0}{d}$ and of an ideal integrator.

\pa For the global model, designed to manage the case of broadband excitations whatever the source location is, we assume that the bidirectional sound field is given by the filtering of the non approximated dipolar sound field by $H(\omega)$. 

We can then write:
\begin{equation}
s_{bidi}(r,\theta,t)=
\frac{c_0}{d} \int_{-\infty}^t 
\bigg(
\frac{1}{r_1}.s(u-\frac{r_1}{c_0})
-\frac{1}{r_2}.s(u-\frac{r_2}{c_0})
\bigg).du.
\label{eq:ChpBidi}
\end{equation}

\pa As said before, one may note that the classical bidirectional directivity is then refound as a limit case: monochromatic excitation of low frequency for a source in \textbf{geometrical} far-field.  

\paragraph{Digital captured bidirectional sound field : }From the expression of the analog bidirectional sound field, we can derive the expression of the digital filter giving the bidirectional sound field from the dipolar digital sound field $s_{dipo}[n]$ given by Eq.~8 if we note $I[z]$ the digital approximation of the analog ideal integrator. With these conventions, the digital filter $H_{bidi}[z]$ to apply to the source signal $s[n]$ to get the digital bidirectional sound field $s_{bidi}[n]$ is then defined by:
\begin{equation}
H_{bidi}[z]=I[z].\frac{c_0}{d}.
\Big(
\frac{1}{r_1}.z^{-D_1}-\frac{1}{r_2}.z^{-D_2}
\Big)
\label{eq:ChpBidi2}
\end{equation}
whose design is illustrated in Fig.~3.

\begin{figure}[h!t]
\begin{center}
\includegraphics[height=4cm]{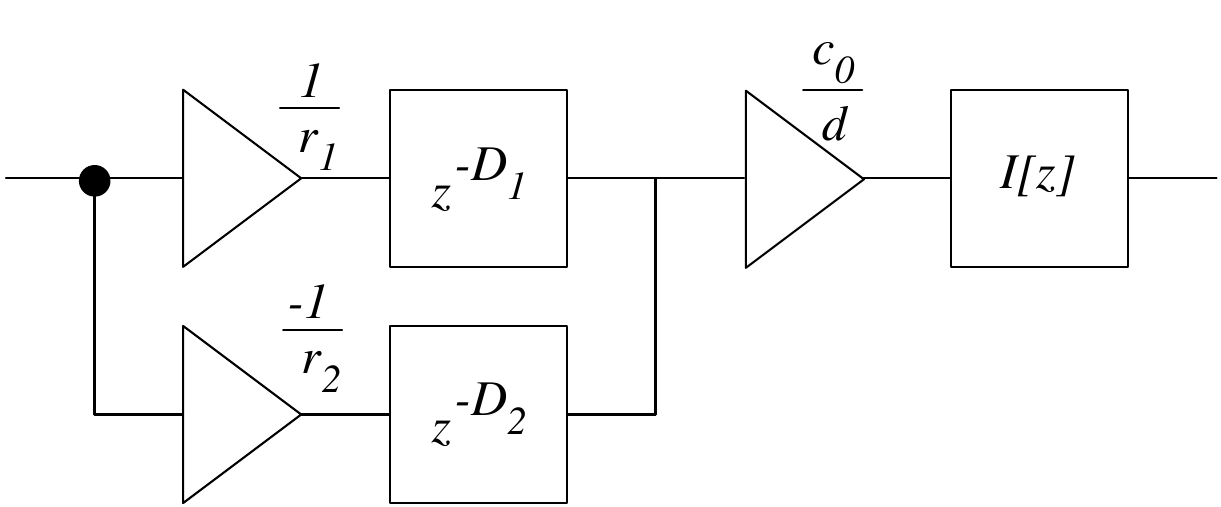}
\end{center}
\caption[3]{Digital filter to apply to the source signal to build a bidirectional acoustical sound field.}
\end{figure}

\pa While several approximations are possible for the digital integrator, we think that the lossy version derived from the definition of the bilinear transform is a good compromise for a real-time application. This version of $I[z]$ is defined by:
\begin{equation}
I[z]=\frac{1}{2.F_s}\frac{1+z^{-1}}{1-g.z^{-1}}
\label{eq:Int}
\end{equation}
where $g$ is a positive real value as close as possible to unity ($g=0.9$ is a value which does not induce instability while working with 16 bits quantization). 

\subsection{Global directional sound field}
\pa So, to synthesize any directional captured sound field, we just need to add the omnidirectional and bidirectional sound fields. As we want to design a realtime application (under Max/MSP), we just need to introduce the definition of the digital filter to apply to any source signal $s[n]$. The transfert function of this global digital filter is:   
\begin{equation}
H[z]=m.\frac{1}{r}.z^{-D_0}
+(1-m).I[z].\frac{c_0}{d}.
\bigg(
\frac{1}{r_1}.z^{-D_1}-\frac{1}{r_2}.z^{-D_2}
\bigg).
\label{eq:Global}
\end{equation}

\pa The implementation of this filter, given by Fig.~4, corresponds to the core of the realtime application. Max/MSP was chosen because we had the possi\-bility to  quickly and rather easily develop a real\-time prototype of this "microphone" using any monophonic sound file as source. 

\begin{figure}[h!t]
\begin{center}
\includegraphics[height=8cm]{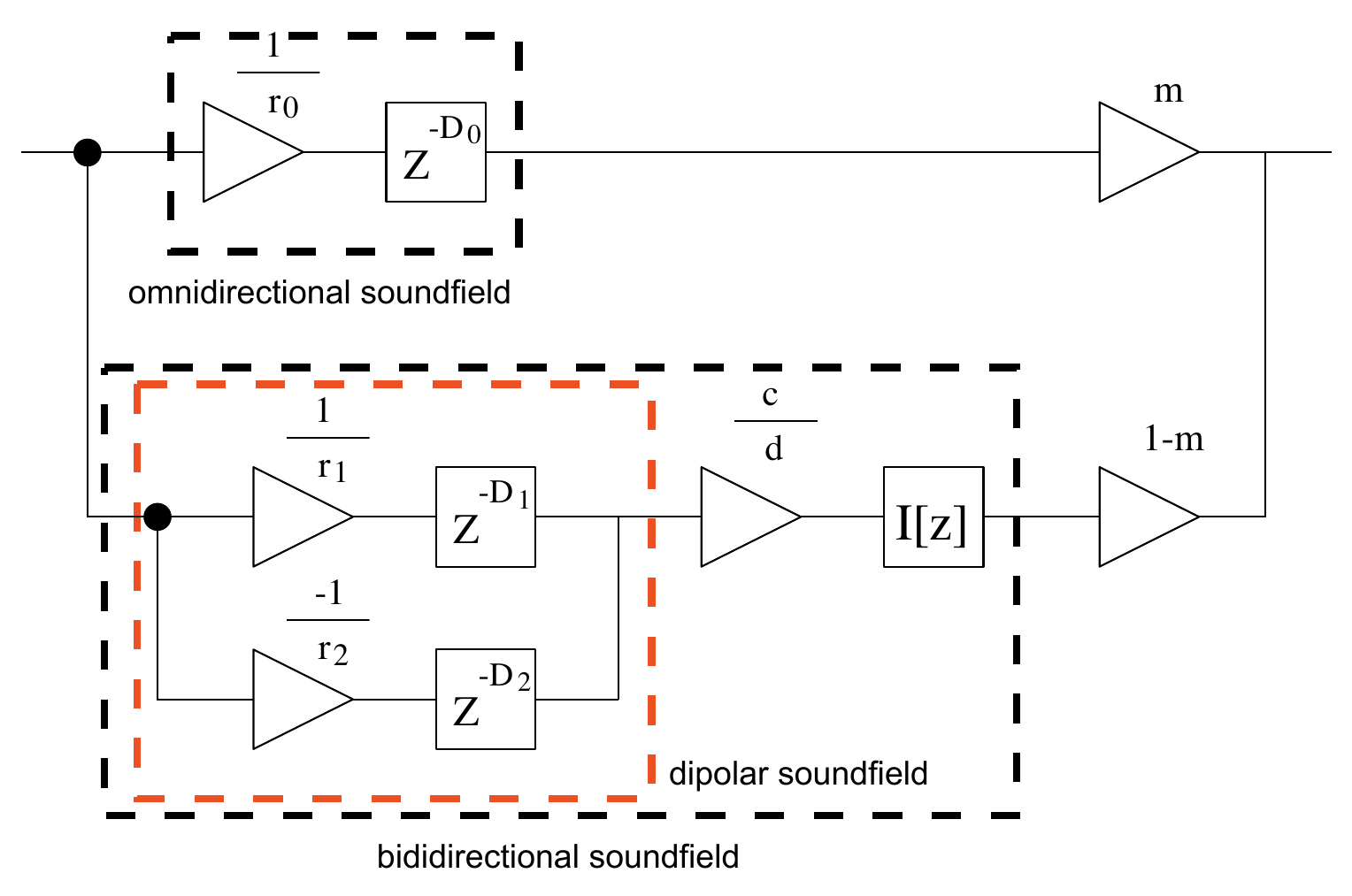}
\end{center}
\caption[4]{Digital filter to apply to the source signal to apply an chosen directional acoustical sound field.}
\end{figure}

\subsection{Directivity digital filter}
\pa It is possible to underline the changes of the "microphone" behavior due to directivity by considering the directivity digital filter, related to the departure from the omnidirectional case. This directivity filter can be derived after factorisation of the omnidirectional component in the global filter $H[z]$ defined by equation (\ref{eq:Global}). So, we first can write:
\begin{equation}
H[z]=\frac{1}{r}.z^{-D_0}.
\big(
m + (1-m).I[z].\frac{c_0}{d}.
\big(
\frac{r}{r_1}.z^{D_0-D_1}-\frac{r}{r_2}.z^{D_0-D_2}
\big)\big)
\end{equation}
and conclude that the directivity filter is finally defined by:
\begin{equation}
H_{dir}[z]=
m + (1-m).I[z].\frac{c_0}{d}.
\big(
\frac{r}{r_1}.z^{D_0-D_1}-\frac{r}{r_2}.z^{D_0-D_2}
\big).
\label{eq:Shift}
\end{equation}

\pa One may note that the directivity filter $H_{dir}[z]$, given by equation (\ref{eq:Shift}), is an extension of the directivity function $D(\theta)$ by a filter which depends on the source location (distance and incidence angle) and on the frequency. Indeed, the filter 
$H_{dir}[z]$ is a  "double" FIR comb filter as it uses a combination of a direct channel and two other delayed ones.

\pa For the discussion, it is useful to note that we have:
\begin{equation}
\begin{split}
H[z] & =\frac{1}{r}.z^{-D_0}.H_{dir}[z],\\
H[z]  & =H_{omni}[z].H_{dir}[z],
\end{split}
\label{eq:Link}
\end{equation}
and that considering source locations so that $r< \dd$ does not make sense. Indeed, the limit case, which is not physically possible when the microphone is effectively present in the considered space, corresponds to the case where the source location coincides with one of the three captation points.

\section{Discussion}
\subsection{Performance of the proposed model}
\pa The complexity of the behavior of the model can not be easily pointed out by "classical plots" (transfert function, monochromatic directivity pattern) and we think the best thing to do is listening the simulation in realtime with various sources and while changing the theoretical directivity and/or the source location. This would be done during the presentation and the associated poster session, and, in the future, we intend to release this tool in a more reliable version than a Max/MSP prototype.

\pa We also believe that the classical plots, transfert functions or monochromatic directivity diagrams, are not adapted to the perception of the quality of the model. Indeed, such analysis use steady monochromatic excitations signals while, in real situations, we use complex sources, for instance music or speech, which both vary with time.

\pa We have been working on the design and programmation of new measurement diagrams more correlated with our perception mainly using a non regular subband analysis (see \cite{Millot:04} for more details on the IDS analysis) and, moreover, using complexe signals as stimuli which may permit to use the same sti\-muli for objective and subjective studies giving the opportunity to study the degree of correlation between objective and subjective results. 

\pa As suggested in \cite{Millot:04}, we consider both global and local analysis which is possible as we can store the subbands sound files if necessary. We have been working on global energy balance as an alternative to the global spectrum and on the way to represent the temporal evolution of this energy balance. We also have been considering the introduction of subband directivity diagrams as an alternative to the classical monochromatic ones. But, some residual programmation bugs still prevent us from getting clean diagrams at the time of this paper writing. We hope this problems would be solved before the AES Convention and would permit us to include some curves and plots in the presentation as the simulations have been recorded (either for the directivity study and the proximity one) and analyzed with the subband analysis: we just miss diagrams with sufficient presentation quality.

\pa A measurements campaign is in progress, with the help of our students of Louis-Lumière. But, as we do not have access to an anechoic chamber, for the moment, measurements are performed in classroom or in a recording studio with music, sinusoids and pink noise. This measurements campaign may be useful to test the protocoles and investigate the potential improvements but, such measurements are also more close to the "everyday" situation met by sound engineers. 

\pa Measurements in an anechoic chamber should be a reality by the end of the year which would permit us to perform measurements in a configuration as similar as possible to the tested one: theoretical free-field study of the acoustical sound field created by a punctual source (a musical monophonic sound file)...

\pa In the following, we review the clues for the extension of this first model and some potential uses of a realtime microphone simulator.

\subsection{Extensions of the simulator}
\pa In the proposed model, we do not physically introduce the microphone as we consider a directional acoustical sound field and then combine three punctual sound fields. A first enhancement may be to introduce the microphone that is to say to take account of its geometry, its mechanical behavior and the transduction process. This can be achieved by using electroacoustical analogies, as Torio \cite{Torio:98} for instance, which may consist in considering the acoustical sound fields as inputs for the model of the microphone whether the membrane is considered as punctual or we introduce its local displacements. 

\pa We need to precise that Millot has found that the electroacoustical analogies do not correspond to the low-frequency limit of acoustical impedances. In fact, these analogies can be directly derived from more general fluids dynamics by assuming particular natures for local flows: adiabatic behavior with instantaneous homogenisation of the pressure in a volume element; unsteady, irrotational and incompressible flow with viscous losses for a port for instance. These derivations will be reported elsewhere as it seem that there could be some serious and general troubles with the impedance formalism.

\pa Another enhancement could be the introduction of the room effect in the model. A first idea can be to introduce only the first reflections and add a diffuse-field model independent of the source location as suggested by Braasch \cite{Braasch:05}. An alternative solution could consist in synthesizing the impulse response  of the room as the combination of the image sources of the source associated with any of the corners of the effective room, and using fast convolution. For a given corner, we can show that we only need to take into account a finite number of image sources, whose number depends on the quantization depth: due to the finite precision of the calculation and to the geometrical attenuation included in the spherical-waves model, only the first image sources would contribute to the sound as the following image sources will be coded as zero.

\pa Introducing the atmospheric attenuation in the model does not seem rather easy as experimental data show it depends on the distance, on the frequency and on other parameters. But, as it seems to be quite tiny for short distances, we think that not considering it in the model should not change significantly the results.

\pa In fact, we think there could be more critical limitations in the proposed model. 

\pa If we sum-up the main assumptions, in this model, we consider a punctual source, the linear combination of three punctual acoustical sound fields and spherical-wave propagation from the source to each of the captation points. So, we can say that we have just kept few  assumptions. 

\pa Considering a punctual source could be a rough assumption as, generally, sources have a spatial extension. But, as we generally miss the information about the sources phenomena, replacing this assumption should be really difficult. Moreover, if we consider the linear assumption, that is to say that each source creates a sound field which linearly interact with each of the sound fields created by the other sources, this assumption could still be used: we just should have to introduce as many other sources as needed  around the first one.

\pa The same discussion remains valid for the captation point. In real configurations, the resultant signal is often assumed to be the linear combination of the sound fields existing in each point of the membrane, and this for both sides of the membrane. But, we then should need to introduce the local displacement of the membrane, which is rather more complicated. Within a first approach, keeping the punctual assumption for the captation seems to be much more reasonable and may permit to feel and/or understand the main behavior of a given microphone.

\pa In fact, the weakest assumption may be the spherical-wave propagation one. Indeed, classical derivation of wave propagation equation assumes that there are absolutely no source in the studied area which does not coincide with our aim to study the proximity effect. So, down to a given distance to the source, unknown in almost if not all the real situations, we can not use the wave model. We (re)find this model failure, for instance, in the spherical-wave propagation because the $\frac{1}{r}$ term (in the omnidirectional signal which can be factorized according to equation 24) becomes greater as we come closer to the source: for the limit case of coincidence of the source and of the study point for the acoustical sound field, this term becomes infinite which does not make sense from a physical point of view as the source magnitude and energy are always finite. 

\pa Some recent theoretical works, also based on experimental observations, performed by Millot and to be reported elsewhere, indicate that the wave assumption could not be as valid as we think. In fact, it appears that we may have to introduce at least three specified areas for acoustical phenomena, in case of the free-field configuration studied in this paper.

\pa In the first area, the source area, one should consider a confined acoustical planar flow which may look like a free jet so it could be described by a 1D laminar flow. Millot assumes that according to the energy level and to the spatial coherence of the source, this first area may present a quite limited extension: the weaker the level is, the smaller is the thickness of this first area. For non coherent spatial vibration of bodies, for instance, this area would have a negligible thickness and one could consider it does not exist at all within the frame of a first approach.

\pa In the second area, the turbulence area, the planar or so-called free jet is transformed in turbulence and vortices because of the viscosity of the medium. In this area, most of the energy is dissipated and the useful description may be a 3D turbulent flow so it would be quite difficult or even impossible to find a mathematical model for this region. To design an omnidirectional loudspeaker or source, as considered in the proposed model, one may search to reduce the thickness of this area and limit it to the very close vicinity of the source. On the contrary, it is possible to keep a quite directive source with lower losses in the turbulent area, by reducing the output velocity of the jet (see Pellerin \cite{Pellerin:04} for a nice example of this process in the case of a loudspeaker).

\pa In the third area, the quiet area, one may think we should find waves propagation as we are rather far away from the sources and as the turbulences are now probably negligible. These assumptions mean that the spatial and temporal variations can still exist, and then produce hearable sound, but that they are far less greater than the ones we can encounter in first area which makes Millot think that we can consider that any particle should follow a quasi straight path (the curvature of the path can be assumed negligible at least locally). This may imply that the particles paths are now distributed over a significant solid angle: the turbulences in the second area have permit to transform a "violent" confined acoustical flow (source area) well described by a planar flow  in a "sweet" distributed one (quiet area) well described by a divergent spherical flow. And, if we consider precisely, within the frame of these assumptions, the mass and momentum conservation laws, we can make the following remarks:
\begin{itemize}
\item if we consider that the particule velocity spatial fluctuations are quite tiny (so almost negligible) and the flow is isentropic, we note that the density and the pressure fluctuations both almost follow a forward spherical wave propagation law but with a propagation speed equal to the particle velocity; 
\item as experimental results show that the propagation speed is equal to the speed of sound $c_0$, the particle velocity magnitude should be nearly equal to $c_0$;
\item without any linearization, the momentum conservation law gives an unsteady, compressible but irrotational flow, with tiny temporal and spatial fluctuations over its path, for any particle.
\end{itemize} 

\pa So, Millot introduces a behavior for the particle velocity (particule velocity nearly equal to $c_0$) which does not coincide with wave propagation and proposes that spherical wave propagation only cor\-responds to the \textbf{theoretical limit case} for the density and pressure fluctuations behavior.

\pa As this alternative theoretical formulation for Acoustics has been found too recently and need to be discussed further, there is no alternative model, for the moment, to replace the divergent spherical wave model for the source. In fact, as maybe often in Acoustics, we only use the theoretical limit case for the third area without taking into account both sources and turbulences areas. 
 
\pa But, to be able to find a good agreement between experiments and numerical simulations, we should try to introduce these two missing areas. Indeed, if we consider, for instance, that in the close vicinity of the source the laminar flow assumption is valid, an alternative explanation for the proxi\-mity effect could be proposed: putting a microphone in a laminar flow should have a similar effect as putting a spoon in the laminar flow escaping from a tap water, that is to say an abrupt stop of the low-frequency flow, also generating high frequency content because of turbulence, for the spoon side facing the flow and a dead-flow area behind the other side of the spoon. For a microphone, and above all for a directional one whose output signal is proportional to the pressure difference between both sides of the membrane, the recorded signal would present a quite important bass and a high frequency content, the so-called bass and high-frequency boosts.

\subsection{Some uses of the simulator}
\pa The global model could be used to build any recording setup simulation using a modular programmation. Thus, we can introduce stereophonic couples as well as multichannel ones to simulate their behavior and/or work on their (pre-)design: such a tool could permit to determine which are the ranges of the microphones locations and/or the microphones arrangements before testing them in real situations. 

\pa Such  simulator could be also constitute an interesting learning tools for sound engineering students to apprehend the behavior of recording systems (from one to more than 5 microphones) as it would permit to switch, in realtime, between different recording setups while keeping the source location constant or to change the source location while listening, in realtime, the related changes. 

\pa We can also suggest that such simulator could be used as another spatialization tool, moreover if automation of the sources motions is included in the application.

\pa Even in the case of a stereophonic version of the simulator, one could, for instance, use the simulator to mix a principal stereophonic couple and additional spot microphones: one quite easy solution would be to simulate a re-recording of the additional spot microphones with the simulator and mix these re-recordings in realtime with the main couple. 

\section{Conclusion}
\pa In this paper we have presented and deeply discussed a theoretical but physical model giving access to an approximation of the acoustical sound field which does not  introduce the real microphone. We have shown how this model constitutes an extension of the classical directivity function and how it can be used to develop a realtime simulator. Some potential uses of such a simulator have been proposed, other may be found. We have also proposed some potential extension for this first step in order to access a better modeling of the acoustical phenomena involved in the captation process. But, we think that this model could be already too much complex to keep or find the feeling of the physics involved in the studied situation and that another extension will not be really easy due to the lack of knowledge of the processes involved.

\pa Some efforts are still needed before being able to propose a potential alternative to the classical measurements tools: transfert functions, directivity patterns, monochromatic measurements... But, the principles of such tools have been discussed: using broadband signals as stimuli and perceptive subband analysis (IDS analysis). And, moreover, this alternative measurement process could be used to study the degree of correlation between objective measurements and subjective assessments using the same stimuli which is nearly impossible with classical measurements tools.

\pa Within the discussion of the model assumptions, some alternative explanations of given acoustical phenomena have been underlined  and we have evoked the possibility of an acoustical formulation without waves! We have other alternative, but no-wave, explanations for other acoustical phenomena.  In fact, for us, such discussions constitute an incitement to reconsider any acoustical phenomena, the classical measurements or theoretical tools, with a different (or new?) look.  By now, these evoked preliminary works have already also "hurt" other fields of the Physics, the ones where waves constitute a main or the principal basement. For instance, some interesting but worrying results have already been found, in Quantum Mechanics: as in Acoustics, it seems possible to propose an alternative formulation without waves which then may induce the fact that we could seriously propose that the duality has no physical sense... 

\pa Reinvestigating Acoustics (and/or Physics) could be more profitable and relevant than one may first think!


\begin{thebibliography}{99}
\bibitem{Williams:04}
M. Williams, {\it Microphone Arrays for Stereo and Multichannel Sound Recording (Vol. 1)}, Editrice Il Rostro, Segrate, 2004

\bibitem{Remy:04}
B. Remy, Laborie A. and S. Montoya, {\it High Spatial Resolution Multichannel Recording}, in: 116th Convention of the Audio Engineering Society, Berlin, Germany, 2004 May 8--11, Preprint 6116. 

\bibitem{Braasch:05}
J. Braasch, {\it A Loudspeaker-based 3D Sound Projection using Virtual Microphone Control (ViMiC)}, in: 118th Convention of the Audio Engineering Society, Barcelona, Spain, 2005 May 28--31, Preprint 6430. 

\bibitem{Martin:05}
G. Martin, {\it A New Microphone Technique For Five-Channel Recording}, in: 118th Convention of the Audio Engineering Society, Barcelona, Spain, 2005 May 28--31, Preprint 6427. 

\bibitem{CCD:04}
A. Carr-Brown, M. Colcy and N. Delatte, {\it Description spatiale du contenu d'enregistrements stéréophoniques}, Master Thesis, ENS Louis-Lumière, France, 2004.

\bibitem{Millot:05}
L. Millot, Gérard Pelé and Mohammed Elliq, {\it Using Perceptive Subbands Analysis to Perform Audio Scenes Cartography}, in: 118th Convention of the Audio Engineering Society, Barcelona, Spain, 2005 May 28--31, Preprint 6340. 


\bibitem{Shure}
Shure, microphone: model Beta 57A (supercardioid), www.shure.com/datasheets/guides-wiredmics\-/beta57a.pdf.

\bibitem{Neumann}
Neumann, microphone: model U89i (5 different directivities), www.neumann.com, cata0127.pdf.

\bibitem{DPA}
DPA, microphone model: 4007 (omnidirectional), www.dpamicrophones.com, DM00622.pdf

\bibitem{Behringer}
Behringer, microphone: model ECM8000 (omnidirectional), www.behringerdownload.de\-/ECM8000/ECM8000\_C\_Specs.pdf

\bibitem{Valette:05}
A. Valette, {\it Modélisation physique de microphone}, Master Thesis, ENS Louis-Lumière, France, 2005. 

\bibitem{Cotterell:02}
P. S. Cotterell, {\it On the theory of second order soundfield microphone}, PhD Thesis, University of Reading, 2002. 

\bibitem{Torio:98}
G. Torio, {\it Understanding the transfert functions of directional condenser microphones in response to different sound sources}, in Microphones and loudspeakers: the ins and the outs of audio, AES, UK Conference 1998. 

\bibitem{Millot:04}
L. Millot, {\it Some clues to build a sound analysis relevant to hearing}, in: 116th Convention of the Audio Engineering Society, Berlin, Germany, 2003 May 8--11, Preprint 6041. 

\bibitem{Pellerin:04}
G. Pellerin, J.-D. Polack and J.-P. Morkerken, {\it Sound source design in the very low frequency domain}, in: 116th Convention of the Audio Engineering Society, Berlin, Germany, 2003 May 8--11, Preprint 6157. 

\end{thebibliography}
\end{document}